\documentstyle[prc,aps]{revtex}
\begin{document}
\draft
\title{ On quasi-bound states in $\eta$- nucleus systems }
\author{J.Kulpa, S. Wycech\thanks{e-mail "wycech@fuw.edu.pl"}}
\address{Soltan Institute for Nuclear Studies,
Warsaw, Poland}
\author{A.M. Green\thanks{e-mail "anthony.green@helsinki.fi"}}
\address{ Department of Physics and Helsinki Institute of Physics, P.O. Box 9,
FIN--00014 University of Helsinki, Finland}

\maketitle
\begin{abstract}
The optical potential  for the $\eta$-meson is calculated. This is based
on a new  
determination of a large $\eta$-N scattering length and
recent experiments of $\eta$ production in two nucleon collisions.
These determine  the  $\eta$ - nuclear  potential well to be about 45 MeV deep
and able to support S, P, D and F quasibound states. Some of the higher
angular momentum states are fairly narrow and offer a  chance of experimental
detection.
\end{abstract}

\newpage
%\narrowtext
\section{Introduction}
\label{intro}

The possibility of $\eta$-nuclear quasi-bound states  was first 
discussed by Haider and Liu  \cite{hai}  and Li et al. \cite{li},
when it was realised that the $\eta$-nucleon interaction is attractive.
Nevertheless, the early ($\pi$,p) experiment looking for these effects was 
not conclusive \cite{chrien}. One possible reason for the difficulty in
the interpretation of those experimental results is that these states are
very broad. This point of view has been
presented in Ref. \cite{oset}, where the widths of $\eta$ nuclear
states were attributed largely to the two nucleon capture mode.
Another possibility, pursued in this letter, is that these states
are bound much more strongly than was generally expected.
If the quasi-bound  states exist, then one may expect them to be narrow,
in few-nucleon systems, and thus be easier to detect there. An indirect
verification  was suggested  by Wilkin \cite{wil},  who interpreted a rapid
slope of  the $ pd \rightarrow \eta ^{3}$He low energy amplitude as a signal 
of a quasibound state. More recently  in a simpler two nucleon case,
very  strong three-body $pp\eta $ correlations were found in measurements
of the $ pp \rightarrow pp \eta$ cross section in the threshold region
\cite{calpp}.

Today, it is possible to revive the question of $\eta$ nuclear
states  on a firm experimental background.
First, definite progress has been
achieved in the understanding of low energy $\eta$-N interactions.
This allows us to calculate the dominant single-nucleon optical potential 
for an $\eta$ in nuclear matter. Second, the measurement of $\eta$ 
production in the nucleon collisions  allows  us to calculate
the absorptive part of the two-nucleon contribution to this optical potential. 

One particular result of the recent studies is the  scattering length
$ a_{\eta N}$ in the $\eta$-N system, which is expected to
be much larger than  was previously estimated. A recent  value of
$ a_{\eta N}=0.75+i0.27 fm $ has been obtained in Refs.\cite{kmat} and 
\cite{svarc}. Also , several other recent models listed  in 
Ref.\cite{kmat} produce  $ \Re \ a_{\eta N}$  much larger than the
$ 0.3 fm $ expected, when the
first calculations of  $\eta$ nuclear  binding were performed.
These models combine interactions in the coupled $\eta$-N ,
$\pi$-N , $\gamma$-N , $\pi \pi $-N channels and use the scattering
data to fix the properties of the $\eta$-N scattering amplitude in the
threshold region.  
In particular a scattering amplitude as large as $0.75 fm $ would generate
an $\eta$-nucleus potential as deep as  $ -80 $ MeV and a very rich
$\eta$-nuclear  spectroscopy. Such an estimate is not realistic, 
however, and it will be shown below that off-shell and nuclear effects
reduce this well depth to about  $ -50 $ MeV. Even so,  it is 
strong enough to bind S, P, D, F states in heavy nuclei. In addition
an absorptive amplitude of $0.27 fm $ would generate an absorptive
potential  $ W_{N} $ as strong as  $ 30 $ MeV in the nuclear center
and the widths
of S states as large as 50 MeV. Such short lived states would certainly
be very difficult to detect. Again, the off-shell and nuclear effects
reduce this absorptive potential strength to about  $ 10 $ MeV.

In addition, experimental cross-sections for the  $ NN \rightarrow NN \eta,$
and $ d \eta$ reactions  determine  that part of the absorptive potential
which describes the inverse $ \eta NN \rightarrow NN $ reaction
rate in nuclei. These reactions are essential to fix an  uncertainty
that has earlier existed in the estimates of the $\eta$ lifetime
in a nucleus. The two-nucleon capture  processes are found to generate
a rather weak absorptive potential  $ W_{NN}$ of $ \approx 3.2 $ MeV
strength at  nuclear matter densities. Furthermore, this mode of
$\eta $ absorption is found to be dominated by the spin triplet NN pairs. 
Altogether,  the optical potential calculated here supports rather a
rich nuclear spectroscopy for the $\eta$ mesons. 
In particular, many  $ 20 $ MeV broad
S-wave states are expected as well as  some less broad 
P, D and F states which may be  generated in medium and heavy nuclei.

This paper contains three main  sections. In section II the
$ \eta$ absorption mechanisms are discussed and the two nucleon
absorption rate is calculated. Section III is devoted to
calculations of the effective $ N \eta$ amplitude in the nuclear
medium. This is done with a  $K$-matrix approach presented in Ref.
\cite{kmat}. Finally in section IV, some nuclear levels of
the $ \eta$ mesons are given.

\section{  Lifetime of the $\eta$ meson in  nuclear matter }
\label{sec2}

The $\eta$ meson lifetime in a nucleus is determined by three basic
reactions
\begin{equation}
\label{d1}
\eta N \rightarrow \pi N
\end{equation}
\begin{equation}
\label{d2}
 \eta (NN)^0 \rightarrow N N
\end{equation}
\begin{equation}
\label{d3} 
\eta (NN)^1 \rightarrow N N.
\end{equation}

Where the superfix denotes the spin of  NN pairs. 
The first process is rather well known. It is described in more detail
in the next section as it turns out to be the dominant one . The other
two reactions (\ref{d2}) and (\ref{d3}) correspond to $\eta$
absorption on two correlated NN pairs in either the 
spin singlet or spin triplet states. The rates for these 
two-nucleon  $\eta$ capture modes have been uncertain for some time and
two extreme views
were expressed. In the first papers \cite{hai},\cite{li} these reactions
were discarded altogether. On the other hand, a model calculation of
Ref. \cite{oset}
indicated large or even dominant effects  from the latter two processes.
Now a purely phenomenological evaluation is possible as the cross
sections for
\begin{equation}
\label{r2}
  p p \rightarrow p p \eta
\end{equation}
\begin{equation}
\label{r3d}
 p n  \rightarrow d \eta
\end{equation}
\begin{equation}
\label{r3} 
 p n  \rightarrow  p n \eta
\end{equation}
have been  measured in the close to threshold region \cite{ber},\cite{chia},
\cite{calpp},\cite{cald},\cite{upps}. The first reaction (\ref{r2})
may be used to calculate the rate of the  process (\ref{d2}), where
the protons are correlated in a spin singlet state. The second reaction 
(\ref{r3d}) gives the rate of $\eta$ absorption on a spin triplet NN pair.

General, but approximate, relationships between scattering cross sections
and the nuclear decay rates are given by the following formulas for the
absorptive optical potentials:
\begin{equation}
\label{w1}
  W_N(r) = \rho(r)[\frac{1}{2}v_{\eta N} \sigma(N\eta \rightarrow \pi N)]
\end{equation}
\begin{equation}
\label{w2}
  W_{NN}^{0,1}(r) = \rho(r)^2[\frac{1}{2}v_{NN}
  \sigma\left(N N\rightarrow (N N)^{0,1}\eta\right)]  \frac{L_1(NN)}{L_2(NN \eta))} 
\end{equation}
\begin{equation}
\label{w2d}
  W_{NN}^{1}(r) = \rho(r)^2[\frac{1}{2}v_{NN}
  \sigma(pn \rightarrow d \eta )] \frac{L_1(NN)}{L_1(d\eta)\psi_d(0)^2},
\end{equation}
where  $\sigma $ are the total cross sections at low energies, 
$v_{NN}$ is the relative velocity in the  NN  system required
to produce slow a $ \eta$, $ v_{\eta N} $ is the $ \eta N$ relative
velocity, $L_1,L_2 $ are the phase space elements defined by 
\begin{eqnarray*} 
L_1(NN)&=& \int \frac{d\bar{p}}{(2\pi)^3}\delta[E-E_{NN}(p)] \, ,\\
L_2(NN\eta)&=&\int\frac{d\bar{p}d\bar{q}}{(2\pi)^6}
\delta[E-E_{NN}(p)-E_{\eta}(q)]  \, ,
\end{eqnarray*}
and  $\psi_d(0) $ is the deuteron wave function at the origin.
These relationships essentially reflect the detailed balance property of
the direct and inverse processes. One correction involving the deuteron
final state is introduced into
the relation (\ref{w2d})  with the following motivation. The meson formation
reactions require high $(\approx 900 MeV/c)$ momentum transfer between the
two nucleons. Therefore, by the uncertainty principle one expects short
distances to be involved and the deuteron to be formed in a coalescent way.
Hence it is the deuteron wave function $\psi_d(0) $ that arises in
Eq.(\ref{w2d}).
The experimental  cross section $ \sigma(pn \rightarrow d \eta )= 93 \mu b$
measured in Ref.\cite{cald} at the excess energy $Q= 56 $ MeV and 
Eq.(\ref{w2d}) generate an absorptive potential of
$ W_{NN}^{1}(0) =4.2 $ MeV strength at the nuclear center. On the other
hand, the  $\sigma(pp\rightarrow  pp \eta ) = 4.9\mu b$ measured
at $Q= 38 $ MeV  \cite{calpp}  yields $ W_{NN}^{0}(0) =1.2 MeV $,
according to Eq.(\ref{w2}). The statistical average
$ W_{NN} =3/4 W_{NN}^{1}+1/4 W_{NN}^{0} $ obtained in this way is small
[ $ W_{NN}(0) =3.4 $ MeV ] and  is of  less
importance than the dominant $ W_{N}(0)\approx 8 $ MeV.

Corrections to relation (\ref{w2}) are necessary, since it is based
on the detailed balance assumption which has clear limitations
in the nuclear medium situation. These corrections  are discussed rather
schematically as the main effect comes from the deuteron relation
(\ref{w2d}). Close to the meson production threshold the inverse reactions
$ NN \rightarrow NN \eta $
rates are strongly enhanced by final state interactions  between the 
two nucleons.
These interactions reflect the proximity of the deuteron or the spin singlet 
virtual state. Such long ranged structures are not formed in the
initial states of $\eta NN \rightarrow NN $  reactions inside  nuclei.
Thus, a correction should be introduced into Eq.(\ref{w2}). It may
be expressed  in terms of a final state wave function $\psi_{NN} $ in
the same way as it was done in Eq.(\ref{w2d}) for the deuteron.
The usefulness of such a parallel between the bound and scattering
states  was demonstrated in Ref.\cite{fald}. Roughly, the enhancement
due to final state  interactions is due to large values of $\psi_{NN}$
at short ranges as compared to the values of the incident wave -- 
a $j_0 $ spherical Bessel function.
In order to average over the interaction range  a simple potential is used
to describe reaction (\ref{r2}). The wave function and an integral of a
transition matrix element over the three-body phase space are calculated
in a way described in
Ref. \cite{app}. It follows that an  enhancement of the  reaction
rate  due to final state $ p p $ interactions  amounts to a factor of 5
at $Q=38$ MeV. The same factor reduces the spin singlet absorptive
potential to an almost negligible value  of  $ W_{NN}^{0}(0) = 0.2 $ MeV. 
The total nuclear matter average is now $ W_{NN}(0) =3.2 $ MeV, and 
it is
this number that is used  for  further calculations in this paper.

A consistency check for this procedure is provided by reaction
(\ref{r3}), 
which has been studied recently  on a deuteron target \cite{upps}. 
For this purpose, the cross section
$\sigma (pn\rightarrow pn \eta ) = 70 \mu b$
found at  $Q= 56 $ MeV  is now used. This particular value of  $Q$ is
chosen  to guarantee the final $S$-wave dominance and
to reduce the $pn$ final state interactions at the same time. Reaction
(\ref{r3}) involves two $ pn $ spin states. Under  the assumption of
statistical proportion  for  these states,  Eq.(\ref{w2})  generates
$ W_{NN}^{1}(0) = 7.7 $ MeV. The effect of $ pn , S=1 $  final state
interactions  reduces this estimate by a factor of
$ \approx 1.5 $. The final $ W_{NN}^{1}(0) $ compares favourably
with the corresponding value obtained from the deuteron.

\section{  Nuclear optical potential for the $\eta$ meson }
\label{sec3}

In this section, a scattering amplitude for the $\eta$-N system
immersed in nuclear matter is studied. This amplitude is based on a 
phenomenological $K$-matrix model of Ref.\cite{kmat}.
In the next stage, it is used to calculate the optical potential for
$\eta$ mesons.

Low energy  $\eta$-N interactions are dominated by two factors:
the  $S(1540)$  resonance and the cusp at  the   $\eta$-N  threshold.
The cusp is seen directly in the  $\pi$-N  channel but it has to
be calculated in the  $\eta$-N channel. This scenario is plotted in
fig.1, which shows the elastic  $\eta$-N scattering amplitude. The
strength of the cusp reflects the  value of  the scattering length.
On the other hand, as the nucleons are bound it is the region just
below the cusp that determines the nuclear optical potential
for $\eta$ mesons. To calculate it we use a model that
contains four basic channels:  $\eta$-N, $\pi$-N ,$\gamma$-N and
$\pi\pi$-N. To simplify the argument it is briefly presented
in a single $\eta$-N channel case. For numerical calculations this
limitation is relaxed, however.

\subsection{  The $\eta$-N scattering amplitude }
\label{sec3a}

To account for the $S(1540)$ resonance the K matrix is  parametrized  as 

\begin{equation}
\label{k2}
 K=  K_{B} + \frac{ff}{E_0-E} ,
\end{equation}
 where   $ E_0 $  is  the  position of the CDD pole related to the
 resonance and $f$  is its coupling  to the  $\eta$-N  channel.
 The first  term  $ K_B $  describes  some potential interactions
 within the channel. The complete model  discussed in  Ref.\cite{kmat}
 contains in general nine phenomenological parameters, those related to
the  $\eta$-N  channel are $f^2=0.225 $, $ E_0= 1541 $ MeV  and
 $ K_B=0.177 fm^{-1} $.

 Now, the T matrix is obtained from  Heitler's  equation
\begin{equation}
\label{k1}
 T=  K +  i K q T,
 \end{equation}
where $ q $ is the relative  $\eta$-N momentum .

A nuclear optical potential for $\eta$ may be given in terms of
this scattering matrix,
if the energy dependence in $ T(E_N+E_{\eta}) $  is accounted for,
and an average over nucleon states is taken. To do this a simple Fermi
gas is  used with a nucleon Fermi level at $-8 $ MeV and nucleon 
energies given by $E_N= U_N+ p_N^2/2M $. The meson energy $E_{\eta}$ is
put equal to zero, and this simulates the conditions  in large
nuclei when the $\eta$ binding energy is close to the bottom of
the mesonic optical potential well.
The  scattering amplitude averaged over the nucleon levels defines an
effective scattering length which is denoted by $ A_{\eta N} $.
The numerical value  $ A_{\eta N}=0.50+i0.092 fm $ is obtained and 
should be compared to the free scattering length
$ a_{\eta N}=0.75+i0.27 fm $.  The difference between these two quantities
reflects the slope of the cusp as well as a rapid fall of the absorptive
amplitude in the subthreshold region.

The optical potential is given by $ A_{\eta N}$ in the standard way
\begin{equation}
\label{pot1} 
  V_N(r) = - \frac{2 \pi}{\mu_{\eta N}} A_{\eta N}  \rho(r) ,
\end{equation}
where  $\rho $ is the nuclear density, $\mu $ is the $\eta$-N reduced mass
and the index $N$ on $ V_N$ indicates the single nucleon origin of this potential. 
The absorptive part of $V_{N}$ given above coincides in the scattering
region with the $ V_{N} $  of  (\ref{w1}). With
$ A_{\eta N} $ obtained above from the  free scattering amplitudes
the optical  potential  becomes $ V_{N}(0)= -53-10 i $ MeV. The depth
of this well  turns out to be much deeper than the pioneering estimates  of
about $ -25 $ MeV obtained in Refs.\cite{hai}, \cite{li} but the absorptive
part is  similar .

For the  nuclear physics of $\eta$ mesons, the energies in the
$\eta$-nucleon system range from about 50 MeV below the  threshold
to 20 MeV above it. This region is dominated  by the $S(1540)$,  
which is supposed to be determined by some short range (quark)
interactions. This inner state is coupled to the channel  states
which change its properties. In the nuclear medium situation 
one needs to distinguish these internal and channel states as
they are differently affected by the nuclear medium. This question
is studied in the next section .

\subsection{ Nuclear medium effects }
\label{sec3b}

 The assumption behind the singularity in our  $ K $ matrix is that
 the  $S(1540)$ is built upon an internal  state $N^*$ ( say a quark
 state ) which is additional to the channel states.
 In this sense it has to appear in any propagators
 in the  intermediate  states  of interest. The energy of this internal
 state $ E_0 $ contains a contribution $ S $ due to its coupling to the
 channel states. This contribution (or energy shift) 
 may be presented  as

\begin{equation}
\label{shi} 
S= f^2 \int d_{3}q \frac{(v(q)/ v(q_o))^2}{(2\pi)^2\mu_{\eta N}}
\frac{1}{E_{N\eta}(q)-E},
\end{equation}
where $ f $ is the  coupling constant from Eq.(\ref{k2}), $ v(q)$ is
a formfactor for the  $ N^*N\eta$ vertex and $q_o$ is an on-shell
momentum.  While the coupling constant is known in this model the
formfactor is not. It has to be very short ranged,
in order to describe the physics involved and also to be consistent
with Eq.(\ref{k2}), where  $ E_0 $ is kept constant. It turns out that, 
for the nuclear physics of interest, the value of $ S $ is irrelevant.
What is relevant is the change $ \Delta S $ due to the nuclear medium,
and this involves only the low  $ q $ region in the integral 
of Eq.(\ref{shi}).
The nuclear shift of the $S(1540)$ is thus practically independent of 
the formfactor. Thus, the main effect of the nuclear medium becomes
the Pauli blocking effect which should be introduced into Eq.(\ref{shi}).  
It pushes the energy of $S(1540)$ upwards. Typical shifts $ \Delta S $ for the nuclear matter Fermi
momentum $ 1.36 fm^{-1} $ are $ 36 $ MeV for a nucleon at the bottom
of the well and $ 19$  MeV  for a nucleon at the Fermi level. In these
calculations the $\eta$ meson has been put to rest on the bottom of its
potential well and the intermediate state
meson excitation energies are described by free kinetic energies.
Since the energy release in the pionic decay  channels  is large, 
the corresponding  nuclear effects due to  $N \pi$ and $N \pi \pi$     
states are insignificant. In principle, the nuclear effects enter 
also into the term $ K_{B} $ of  Eq.(\ref{k2}). However, this term is 
very small in this model and these corrections are negligible. 
Readers interested in the details of such calculations 
are referred to  Ref.~\cite{sta}, where similar shifts are 
discussed in the context of K mesons. 
In practice, this model of $S(1540)$ and its nuclear behavior 
are close to the description of  the $\Lambda (1405)$ 
used in Ref.~\cite{eis}, 
where all the nuclear effects enter via the $ S $ of Eq.(\ref{shi}).
The nuclear shift $ \Delta S $ has  a surprisingly  small 
effect on the depth of  the  $\eta$ meson optical potential 
well but it reduces the absorption strength. When included into the  scattering
matrix  this shift produces an effective scattering length in the nuclear 
medium  $ A_{\eta N}=0.45 +i0.068  fm $ and the optical potential depth 
$ V_{N}(0) = -47-7.1 i $ MeV. This value is used in further calculations. 

Another but related point to discuss is an effect of the nuclear absorption
on the nuclear properties of  the $ S(1540) $. The integral in Eq.(\ref{shi})
is normalized in such a way that its imaginary part yields a half width 
for  this state.  This width may be changed by a collision damping 
process    
i.e.  the  $ N^{*} N \rightarrow  N N $  reaction. In the previous 
section this process was related  to the experimental  
$ N N \rightarrow N  N^{*} \rightarrow  N N \eta $  cross section 
and this relation has generated the 
two nucleon absorptive potential $ W_{NN}$. 
One might be tempted to include this 
process also into Eq.(\ref{shi}) introducing  complex energies 
$ E $ and $E_{N\eta}$. 
Corrections to the optical potential  
obtained in this way are found to be small 
both for the real and the absorptive parts.  
However, as such a method is model 
dependent these corrections have been dropped altogether.

\section{RESULTS}

The basic difference between this and previous 
calculations is that the  $\eta$-N  model used  here 
produces the $\eta$- nuclear optical potential 
well to be  much deeper than those due to other models. 
The nuclear states of eta mesons are thus bound much more 
strongly. On the other hand, the  
absorptive part of this potential is 
comparable to results obtained elsewhere.   
The effect of two nucleon capture modes 
calculated directly from the scattering cross sections  
is rather moderate. It enlarges the level widths,  in particular those 
of the S  states,  by about $ 5 $ MeV. 
On the other hand,  widths of higher angular momentum states localized at the 
nuclear surface  are less affected, since they involve  $ \rho^2 $  terms. 
Some quasibound state energies and half widths in are given in  Table I . 
These were calculated with the nuclear densities following the electric 
charge profiles. In addition 
to those states there may arise resonances generated 
by the centrifugal barrier. For example,  a resonance in  a G wave 
is likely to be formed in this way in Pb.

\begin{table}
\caption{
Energies and half-widths of eta states in  nuclei,  $[MeV] $. }
\begin{tabular}{lcccc}
   Nucleus          & $S         $     &   $P  $            &  $D  $            & $F $  \\ \hline 
    $ ^{16}$O        & --17.2 -- i 6.7   &      --           &    --           & --            \\
    $^{40}$Ar        & --27.2 -- i 8.6   &  --11.2  -- i 6.6  &                & --            \\
   $^{208}$Pb        & --39.2 -- i 10.1  &  --31.5  -- i 9.6  &  --22.6 -- i 9.0 & --12.3
-- i 8.1   \\
    $^{208}$Pb       
& --19.4 -- i 8.6   &  --8.0   -- i 7.3  &  --             & --             \\
\end{tabular}
\label{table1}
\end{table}

The mesons appear to be strongly bound even in light nuclei, in particular
the S state obtained in  
oxygen is bound  by about $15 $ MeV more than the one predicted 
in Ref.\cite{hai}.

A similar calculation has been performed in very light nuclei. 
In particular,
$S$ wave bound states were  obtained in $^{4}$He  and $ ^{6}$He with
the complex energies of $ -8.8 - i 7.4 $ and $ -17.2 - i 9.5 $ MeV. 
The latter system is of interest in the context of  experiments planned 
at the Brookhaven Laboratory \cite{nefkens},  where it may be observed as
a final state. The energy given above is based on the structure and radius of $ ^{6}$He found  in  Ref. \cite{alk}.

Another effect of the strong attraction is that high angular momentum bound
states arise. Some of those upper states may  depended on details like:
neutron density distributions, the
distribution of nuclear momenta in the surface region and the range of
$\eta $N  forces
which have not been discussed here. These effects are of secondary importance
and should be discussed only if there is a significant
experimental progress in this field.  The upper
bound states  are slightly narrower and in some cases may be  separated from the region of 
strongly overlapping lower states.  
The experimental search  for those states may turn out to be easier.

\vspace{0.5cm} 

\noindent Acknowledgements.
We wish to thank Joanna Stepaniak for many stimulating discussions.  
S.W. is grateful to the Helsinki Institute of Physics  for hospitality and
financial support. Support from KBN Grant No 2P0B3 048 12 is  also 
acknowledged.

\begin{figure}
\caption[F1] 
{The elastic $\eta$-N scattering amplitude in $fermi$ units
plotted against the C.M. energy Ecm  in MeV. Real part - continuous line, 
absorptive part - dashed line.}
\label{fig1}
\end{figure}
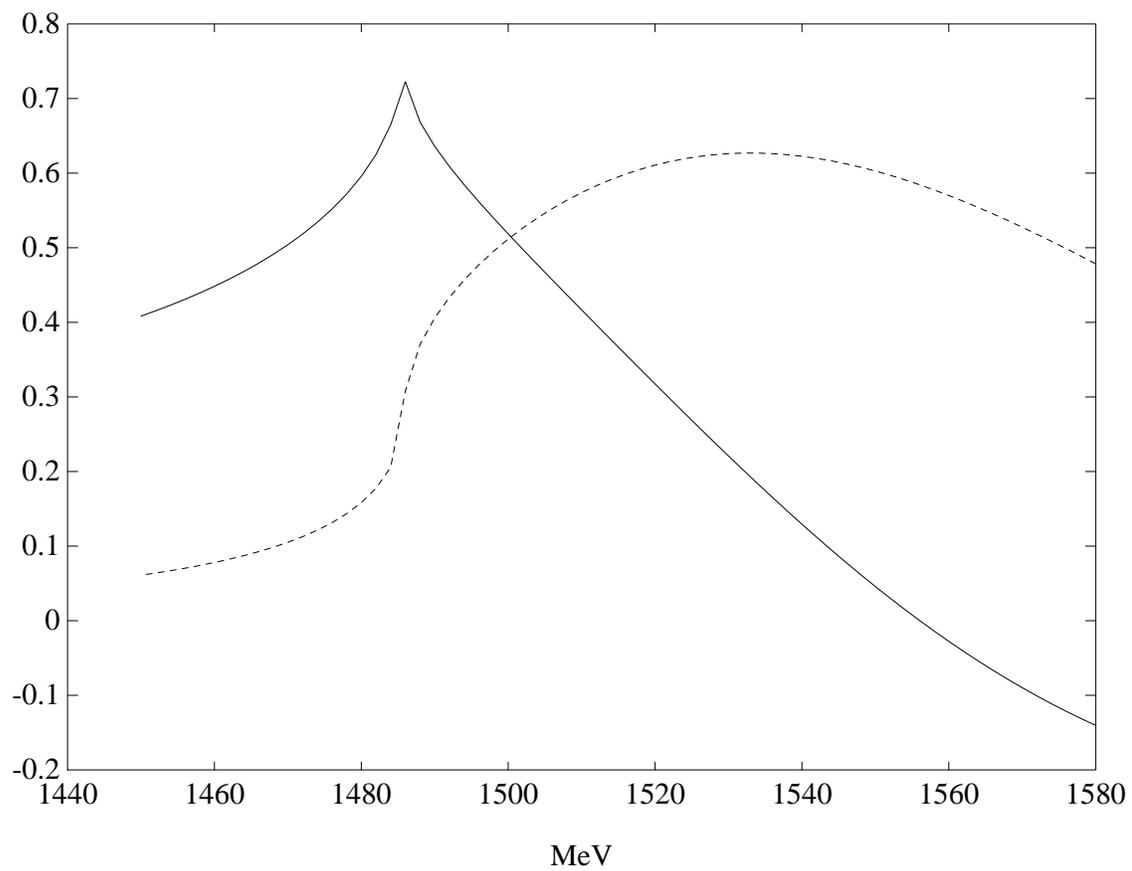
\end{document}